\documentclass[prb,aps,showpacs,preprint]{revtex4-1}
\usepackage{epsf,enumerate}
\usepackage[dvips]{color}
\usepackage{graphicx}
\usepackage{bm}

\renewcommand\Re{\mathrm{Re\,}}

\definecolor{green}{rgb}{0,0.5,0}

\def\be{\begin{equation}}
\def\ee{\end{equation}}

\begin{document}

\title{Morphological Transitions in Nanoscale Patterns Produced by Concurrent Ion Sputtering and Impurity Co-deposition}

\author{R. Mark Bradley}
\affiliation{Department of Physics, Colorado State University, Fort
Collins, CO 80523, USA}

\date{\today}

\begin{abstract}

We modify the theory of nanoscale patterns produced by ion bombardment with concurrent impurity deposition 
to take into account the effect that the near-surface impurities have on the collision cascades. 
As the impurity concentration is increased, the resulting theory successively yields a flat surface, a rippled surface with its wavevector along the projected direction of ion incidence, and a rippled surface with its wavevector rotated by $90^\circ$.  Exactly the same morphological transitions were observed in recent experiments in which silicon was bombarded with an argon ion beam and gold was co-deposited [B.~Moon {\it et al.}~arXiv:1601.02534 (2016)].  

\end{abstract}


\date{\today}

\maketitle

\section{Introduction}
\label{sec:introduction}

Bombarding a solid surface with a broad ion beam can produce a remarkable variety of self-assembled 
nanoscale patterns, including periodic height modulations and 
mounds arranged in hexagonal arrays of surprising regularity.~\cite{Ziberi09,Facsko99,Frost00,Munoz-Garcia14}
The spontaneous formation of these patterns is
not just fascinating in its own right: Ion bombardment has the
potential to become a cost-effective method to rapidly fabricate
large-area nanostructures at length scales beyond the limits
of conventional optical lithography.

Over the past decade, an abundance of experimental work has established that the deposition of impurities during ion bombardment of an elemental material has a profound effect on the nanoscale patterns that develop on its surface.\cite{Ozaydin05,Ozaydin08,Hofsass08,Sanchez-Garcia08,Ozaydin-Ince09,Macko10,Zhang10,Zhou11,Cornejo11,Zhang11,Macko11,Hofsass12,Zhang12,Macko12,
Redondo-Cubero12,Hofsass13,Bhattacharjee13,Khanbabaee14,Engler14,Gago14,Ko15,Vayalil15}  The pioneering experimental work in the field was carried out by Ozaydin {\it et al.},\cite{Ozaydin05,Ozaydin08,Ozaydin-Ince09} who bombarded a silicon sample at normal incidence in ultrahigh vacuum and found that it remained flat.  In contrast, if a trace amount of molybdenum atoms was deposited during the bombardment, a disordered array of nanodots formed on the surface.

Zhang, Br\"otzmann and Hofs\"ass subsequently observed that if silicon is subjected to normal-incidence ion bombardment with concurrent oblique-incidence deposition of iron atoms,
surface ripples develop if the iron flux is sufficiently high.\cite{Zhang11}  The ripple wavevector 
was parallel to the surface projection of the incidence direction of the iron impurities.
In follow-up work, these experiments were repeated, but with a range of metals replacing Fe, including 
Ni, Mo, W, and Au.\cite{Hofsass12}  The most pronounced tendency to form ripples was observed for the metals Fe, Ni, Mo, and W, which form Si-rich disilicides
with the stoichiometry MeSi$_2$.  Gold, on the other hand, has no stable silicides and no patterns were observed, even for high Au surface coverages.  These results suggest that 
compound formation can play a key role in determining whether or not surface ripples develop during ion bombardment with co-deposition of impurities.

Inspired by the work of Zhang, Br\"otzmann and Hofs\"ass, we demonstrated that if impurities are
deposited obliquely on a solid surface during normal-incidence ion bombardment, an instability can occur that results purely from the interaction between the topography of the surface and the surface 
layer in which impurities are present.\cite{Bradley12} This instability only occurs if the sputter yields of the target material and the impurities differ.  It leads to the formation of a surface ripple with its wavevector along the projection of the impurity beam onto the sample surface, as observed by Zhang, Br\"otzmann and Hofs\"ass.  In later work, the effect of compound formation was incorporated into the theory.\cite{Bradley13}

Intriguing recent experiments by Moon {\it et al.}~present a challenge to the prevailing view of ion sputtering with concurrent impurity deposition.\cite{Moon16}  In these experiments, a silicon surface was bombarded with an argon ion beam and gold impurities were simultaneously deposited.
Both the argon ions and gold atoms were obliquely incident on the surface.
Gold was chosen as the impurity because it has no stable silicides.  The angle of ion incidence was selected in an especially clever fashion: for the chosen angle, the sputter yields of gold and silicon are equal.
Despite this, ripples with their wavevector parallel to the projected impurity incidence direction formed on the solid surface if the gold concentration was sufficiently high.  What is more, for still higher gold concentrations, the ripple wavevector rotated by $90^\circ$.  Finally, for sufficiently low gold concentrations, the surface remained flat.

The experimental results of Moon {\it et al.}~cannot be explained by the theory advanced in Ref.~\onlinecite{Bradley12} because the silicon and gold had the same sputter yield in those experiments.  Additionally, the theory in Ref.~\onlinecite{Bradley12} does not yield ripples with their wavevector perpendicular to the projected impurity incidence direction for any choice of parameter values.  

In this paper, we modify the existing theory of ion bombardment with concurrent impurity deposition 
to take into account the effect that the near-surface impurities have on the collision cascades. 
This requires that the Sigmund model of ion sputtering\cite{Sigmund73} be generalized.
Once this crucial modification has been made, the theory can produce precisely the same phenomena as Moon {\it et al.}~observed.  In particular, as the impurity concentration is increased, the theory can successively yield a flat surface, a rippled surface with its wavevector along the projected direction of ion incidence, and a rippled surface with its wavevector rotated by $90^\circ$. 

To make our theory as widely applicable as possible, we will not restrict our attention to Ar-ion bombardment of Si with Au co-deposition.  Instead, we will study a sample initially composed of entirely of atomic species $B$ that is subjected to bombardment with a noble gas ion beam and to concurrent deposition of atoms of species $A$.  We will make several assumptions that are motivated by the experiments of Moon {\it et al.} --- for example, we will assume $A$ and $B$ do not react chemically to form a stable compound.  However, a number of the parameters in the theory are not currently known for a silicon target with gold co-deposition.  In some cases, we will make simplifying assumptions about these parameters that may or may not be valid for the experiments of Moon {\it et al.}  Our goal is to demonstrate in the simplest possible context that the phenomena observed by Moon {\it et al.} can arise from the existing theory once it has been suitably modified, not to produce a theory that quantitatively reproduces all of their measurements.  Parameter values computed using atomistic simulations or measured experimentally will be needed before a theory that precisely matches the results of Moon {\it et al.}~can be developed.

This paper is organized as follows.  In Sec.~II, the Sigmund theory of sputtering is modified so that 
the effect of the near-surface composition on the collision cascades is taken into account.  The resulting equations of motion are derived in Sec.~III and are analyzed in Sec.~IV.  We discuss our findings in Sec.~V and close by summarizing them in Sec.~VI.

\section{Generalized Sigmund Model}
\label{sec:general}

Consider an ion of energy $\epsilon$ that impinges on the surface of an elemental solid at the point $Q$.  By shifting the location
of the origin $O$ if necessary, we can arrange for $O$ and $Q$ to coincide.  The height of the surface above the $x-y$ plane, $h=h(x,y,t)$, then vanishes
for $x=y=0$.
  
In the Sigmund theory of sputtering,\cite{Sigmund73} the rate with which material is sputtered from a point on the solid surface
is proportional to the power $P$ deposited there by the random slowing down of ions.  The average
energy density deposited at a point $(x,y,z)$ within the solid by an ion which travels along the $z$ axis until striking the
surface is taken to be
\be 
E(x,y,z) = \frac{\epsilon}{(2\pi)^{3/2}\alpha\beta^2} \exp\left(-\frac{(z+a)^2}{2\alpha^2} - \frac{x^2+y^2}{2\beta^2}\right ).
\label{energy_deposited}
\ee
Here $a$ is the average depth of energy deposition and $\alpha$ and $\beta$ are the longitudinal and transverse
straggling lengths, respectively.  The contours of equal energy deposition are ellipsoids of revolution 
centered at the point $-a\bm{\hat z}$ with the $z$-axis as their axis of symmetry.  

If the angle of incidence $\theta$ is positive rather than zero, the direction of the incident beam is $-\bm{\hat e}$, where $\bm{\hat e}\equiv \bm{\hat x}\sin\theta + \bm{\hat z}\cos\theta$.  In the Sigmund model, the average distribution of deposited energy is obtained
by rotating the distribution (\ref{energy_deposited}) through the angle $\theta$ about the $y$-axis.  Explicitly, the density of deposited energy at an arbitrary point $\bm{r}$ within the solid is given by 
\be 
E(\bm{r}) = \frac{\epsilon}{(2\pi)^{3/2}\alpha\beta^2} \exp\left(-\frac{\rho_\parallel^2}{2\alpha^2} - \frac{\rho_\perp^2}{2\beta^2}\right ),
\label{distn}
\ee
where
\be 
\rho_\parallel = a + x\sin\theta + z\cos\theta
\label{rhopara}
\ee
and
\be 
\rho_\perp = \left[(x\cos\theta - z\sin\theta)^2 + y^2\right]^{1/2},
\label{rhoperp}
\ee
as shown in Ref.~\onlinecite{Bradley11exact}.

The Sigmund model of ion sputtering of an elemental material has been extended so that it applies if atoms of two different atomic species $A$ and $B$ are present in a layer at the surface of the solid.\cite{Shenoy07,Bradley10,Shipman11,Bradley11imp,Bradley12ASS}
In the extended model, it is assumed that the sputtered fluxes
of the two species are proportional to $P$.  Let those fluxes be denoted by $F_A$ and $F_B$, respectively.  One also makes the reasonable supposition that $F_A$ is proportional to 
the concentration of $A$ atoms at the surface, which will be denoted by $c_s$. Thus, 
\be 
F_A = \lambda_A c_s P,
\label{F_A}
\ee
and similarly
\be 
F_B = \lambda_B (1-c_s) P,
\label{F_B}
\ee
where the constants of proportionality $\lambda_A$ and $\lambda_B$ are positive.  If the atoms of species $A$ (species $B$)
are preferentially sputtered, then $\lambda_A$ is greater than (less than) $\lambda_B$. 

In the theory of Shenoy {\it et al.}\cite{Shenoy07} and in subsequent work,\cite{Bradley10,Shipman11,Bradley11imp,Bradley12ASS} it was assumed that a change in the target's composition in the near-surface region has no effect on the collision cascades.  This assumption is unlikely to be valid except in unusual circumstances.  Here we will take a first step beyond this crude approximation by making an additional modification to the Sigmund model.  Let $E_A(\bm{r})$ be the average density of deposited energy in a target composed entirely of species $A$ and define $E_B(\bm{r})$ analogously.  
We take $E_i(\bm{r})$ to be given by Eq.~(\ref{distn}), but with $a$, $\alpha$ and $\beta$ replaced by $a_i$, $\alpha_i$ and $\beta_i$, where $i=A$ or $B$.  In general, $a_A$, $\alpha_A$ and $\beta_A$ differ from $a_B$, $\alpha_B$ and $\beta_B$ because collision cascades are material-dependent.  If atoms of both atomic species are present in a surface layer, we will assume that the density of deposited energy is given by
\be
E(\bm{r})=c_s(\bm{r}) E_A(\bm{r})+[1-c_s(\bm{r})]E_B(\bm{r}).
\label{interpolation}
\ee
This linear interpolation between the limits $c_s=0$ and $c_s=1$ is likely an imperfect approximation, but it is better than simply neglecting the dependence of $E$ on target composition as has been done in previous work.

\section{Derivation of the Equations of Motion}
\label{sec:eoms}

Consider an elemental solid consisting of atoms of species $B$.  Initially, the solid occupies the region with $z\le 0$ and has a planar
surface.  The solid is now subjected to ion bombardment with concurrent deposition of atoms of species $A$, as shown in Fig.~1.  Above the surface of the solid, the ion flux is $-J\bm{\hat e}$.  The impurity flux, on the other hand, is $-J_d\bm{\hat e}_d$, where $\bm{\hat e}_d = \bm{\hat x}\sin\theta_d  + \bm{\hat z}\cos\theta_d$ and $\theta_d$ is the angle of incidence of the impurities.  We will restrict our attention to positive values of
$\theta$, but the full range of angles $-\pi/2< \theta_d < \pi/2$ will be considered.  

Motivated by the experiments of Moon {\it et al.}, we assume that 
species $A$ and $B$ do not react chemically and that the energy of the incident impurity atoms is low enough that they do not sputter atoms from the surface of the solid.  We take also the incident flux of impurities $J_d$ to be small enough compared to the ion flux $J$
that the net result of the concurrent bombardment and deposition is erosion of the solid.  
For simplicity, we take the atomic volume $\Omega$ to be the same for both species and assume that phase separation does not occur.

\vspace{15pt}

\centerline{\includegraphics[width=3.5in]{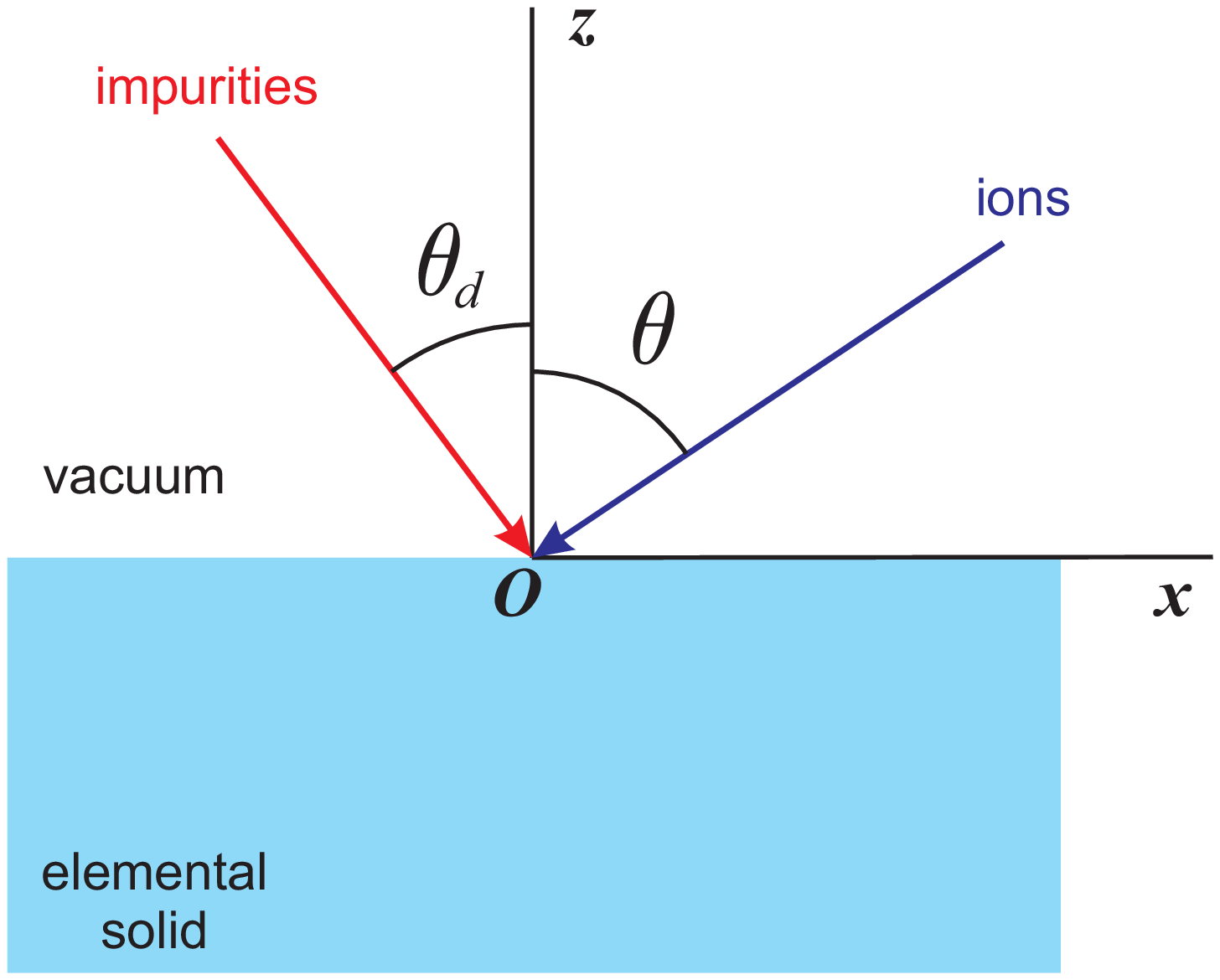}}
\begin{figure}[h]
\caption{(Color online) The initial state of the solid, showing the directions of the incident ions and impurities.  $\theta_d$ is negative in the figure.}
\label{fig1}
\end{figure}

As time passes, a surface layer develops in which atoms of species $A$ are present.
A steady state is eventually reached in which the concentration of $A$ atoms at the surface has a constant value $c_{s,0}$ and the solid is eroded at a constant rate, 
i.e., the surface height $h=h_0-v_0t$, where $h_0$ and $v_0>0$ are constants.

Suppose that the planar surface is now disturbed slightly.  The surface height 
\be 
h=h_0-v_0t+u 
\ee
and the concentration of $A$ atoms at the surface 
\be 
c_s=c_{s,0}+\phi 
\ee
are then functions
of $x$, $y$ and $t$.  The deviations of the surface height and the surface composition
from their steady-state values ($u$ and $\phi$, respectively) are small.  We will work to first order in these quantities.

As shown in detail in Ref.~\onlinecite{Bradley11imp}, the equations of motion are 
\be 
{{\partial h}\over {\partial t}} = -\Omega(F_A+F_B-F_d+{\bm\nabla}\cdot {\bm J}_A + {\bm\nabla}\cdot {\bm J}_B),
\label{h_eom}
\ee
and
\be 
\Delta{{\partial c_{s}}\over {\partial t}}= -\Omega(F_A - F_d +{\bm\nabla}\cdot {\bm J}_A).
\label{c_eom}
\ee
Here $\Delta$ is the thickness of the surface layer of altered composition, 
\be 
F_d = J_d (\cos\theta_d - u_x\sin\theta_d)
\label{F_d}
\ee
is the deposited flux of $A$ atoms, and $\bm{J}_i$ denotes the surface current of atoms of species $i$, where $i=A$ or $B$.

Suppose that the surface height $h$ varies slowly with position.  If the target were composed of an elemental material, then the power deposited per unit surface area would be given by
\be 
P = P_0 - \gamma u_x - \nu_1 u_{xx} - \nu_2 u_{yy},
\label{power_elemental}
\ee
where the subscripts $x$ and $y$ denote partial derivatives and we have dropped terms that involve three or more spatial derivatives.\cite{Bradley88,Motta14}  Explicit expressions for the coefficients $P_0$, $\gamma$, $\nu_1$ and $\nu_2$ may be found in Ref.~\onlinecite{Motta14}.  The analog of Eq.~(\ref{power_elemental}) for our target material with both $A$ and $B$ atoms present in a surface layer can be readily obtained.  Using Eq.~(\ref{interpolation}), we find that
\be
P= c_s P_A + (1-c_{s,0})P_B,
\ee
where $P_i$ is the power per unit surface area that would be deposited if the target were composed entirely of species $i$ and $i=A$ or $B$.  Hence
\be
P=c_s(P_{0,A} - \gamma_A u_x - \nu_{1,A} u_{xx} - \nu_{2,A} u_{yy})
+(1-c_s)(P_{0,B} - \gamma_B u_x - \nu_{1,B} u_{xx} - \nu_{2,B} u_{yy}),
\ee
where $P_{0,i}$, $\gamma_i$, $\nu_{1,i}$ and $\nu_{2,i}$ are the values of $P_0$, $\gamma$, $\nu_1$ and $\nu_2$ for a target made up exclusively of atoms of species $i$. 
The sputtered fluxes of the two atomic species may now be obtained from Eqs.~(\ref{F_A}) and (\ref{F_B}).

Ion bombardment can induce viscous flow in a layer near the surface of a solid.\cite{Umbach01}  In the case of a silicon target maintained at room temperature (as in the experiments of Moon {\it et al.}\cite{Moon16}), these currents are large compared to those produced by surface diffusion, and we will neglect the latter.
The resulting surface current of species $i$ is 
\be 
{\bm J}_i^{(v)} = K c_i {\bm \nabla}\nabla^2 h ,
\label{surface_current}
\ee
where $c_A=c_s$, $c_B=1-c_s$ and $i=A$ or $B$.\cite{Bradley12}  Here $K= \gamma_s\Delta^3/(3\Omega\eta)$, where $\gamma_s$ is the surface tension
and $\eta$ is the radiation-enhanced viscosity.\cite{Umbach01,Orchard63}

Momentum transfer from the incident ions to atoms at the surface also produces surface atomic currents of species $A$ and $B$, i.e., mass redistribution.\cite{Carter96,Moseler05,Davidovitch07}
Adopting the same assumptions as in Ref.~\onlinecite{Motta14}, we obtain the following expressions for those currents:
\begin{eqnarray} 
{\bm J}_A^{(m)} = &-J\mu_A\biggl\{  \frac{1}{2}c_{s,0}\sin(2\theta)({\bm\hat{\bm x}}+u_x{\bm\hat{\bm z}}) + \frac{1}{2}\phi\sin(2\theta){\bm\hat{\bm x}}\nonumber\\
&+ c_{s,0}[u_x \cos(2\theta){\bm\hat{\bm x}} + u_y\cos^2\theta {\bm\hat{\bm y}}]\biggr\}
\label{mom_transf_A}
\end{eqnarray}
and
\begin{eqnarray}  
{\bm J}_B^{(m)} = &-J\mu_B\biggl\{  \frac{1}{2}(1-c_{s,0})\sin(2\theta)({\bm\hat{\bm x}}+u_x{\bm\hat{\bm z}}) - \frac{1}{2}\phi\sin(2\theta){\bm\hat{\bm x}}\nonumber\\
&+ (1-c_{s,0})[u_x \cos(2\theta){\bm\hat{\bm x}} + u_y\cos^2\theta {\bm\hat{\bm y}}]\biggr\}.
\label{mom_transf_B}
\end{eqnarray} 
Here $\mu_i$ is a positive constant that characterizes the ease with which species
$i$ is driven over the surface for $i=A$ and $B$. The values of these constants are not known for 
a silicon sample with gold present in a layer at the surface of the solid.
For the sake of simplicity, we will assume that 
$\mu_A=\mu_B\equiv \mu$.  
The total surface current of species $i$ is ${\bm J}_i = {\bm J}^{(v)}_i + {\bm J}^{(m)}_i$.

To zeroth order in $u$ and $\phi$, the equations of motion (\ref{h_eom}) and (\ref{c_eom}) 
are 
\be
\frac{v_0}{\Omega} = [c_{s,0}\lambda_A + (1-c_{s,0})\lambda_B][c_{s,0}P_{0,A} + (1-c_{s,0})P_{0,B}] - J_d \cos\theta_d
\label{zeroth_1}
\ee
and
\be
c_{s,0}\lambda_A [c_{s,0}P_{0,A} + (1-c_{s,0})P_{0,B}] = J_d \cos\theta_d.
\label{zeroth_2}
\ee
Equations (\ref{zeroth_1}) and (\ref{zeroth_2}) respectively give the erosion velocity $v_0$ and the surface concentration $c_{s,0}$ in the steady state. 

Because Moon {\it et al.}~chose the angle of ion incidence so that the sputter yields of silicon and gold coincided, we will require that $v_0$ have the same value for $c_{s,0}=0$ and $c_{s,0}=1$. We will actually go further and require that $v_0$ be independent of $c_{s,0}$ for $0\le c_{s,0} \le 1$ so that the surface instability cannot be caused by the dependence of the sputter yield on the surface composition as it is in Ref.~\onlinecite{Bradley12}.  A straightforward analysis then shows that
\be
\lambda_A=\lambda_B\equiv \lambda
\ee
and 
\be 
P_{0,A}=P_{0,B}\equiv P_0.
\ee

Retaining terms of first order in $u$ and $\phi$, the equations of motion (\ref{h_eom}) and (\ref{c_eom}) reduce to
\be 
u_t =  C_1 u_{xx} + C_2 u_{yy} - D \nabla^2\nabla^2 u  + \alpha u_x
\label{dim_u_eom}
\ee
and
\be 
\phi_t = -A'\phi + C'_1 u_{xx} + C'_2 u_{yy} - D' \nabla^2\nabla^2 u  +\alpha' u_x+ \beta' \phi_x,
\label{dim_phi_eom}
\ee
where
\begin{eqnarray} 
C_1 &=& \Omega\lambda[c_{s,0}\nu_{1,A} + (1-c_{s,0})\nu_{1,B}]+\Omega\mu J\cos(2\theta) ,\label{C_1}\\
C_2 &=& \Omega\lambda[c_{s,0}\nu_{2,A} + (1-c_{s,0})\nu_{2,B}]+\Omega\mu J\cos^2\theta ,\label{C_2}\\
D &=& \Omega K,\\
\alpha &=& \Omega\lambda[c_{s,0}\gamma_A + (1-c_{s,0})\gamma_B]-\Omega J_d\sin\theta_d,\\
A' &=& \Omega \lambda P_0/\Delta,\\
C'_1 &=& \Omega c_{s,0}\{\lambda[c_{s,0}\nu_{1,A} + (1-c_{s,0})\nu_{1,B}]+\mu J\cos(2\theta)\}/\Delta,\\
C'_2 &=& \Omega c_{s,0}\{\lambda[c_{s,0}\nu_{2,A} + (1-c_{s,0})\nu_{2,B}]+\mu J\cos^2\theta\}/\Delta,\\
D' &=& \Omega c_{s,0} K/\Delta,\\
\alpha' &=& \Omega\{\lambda c_{s,0}[c_{s,0}\gamma_A+(1-c_{s,0})\gamma_B]-J_d\sin\theta_d\}/\Delta,
\end{eqnarray}
and
\be 
\hspace{-20.1em} \beta' = \frac{\Omega J\mu}{2\Delta} \sin(2\theta).
\ee
Note that with the assumptions and approximations that we have made have, $u$ evolves independently of $\phi$.  This greatly simplifies the analysis of the equations of motion. 

\section{Analysis of the Equations of Motion}
\label{sec:analysis}

To probe the stability of the unperturbed steady-state solution $u=\phi=0$,
we seek solutions to the linearized equations of motion~(\ref{dim_u_eom}) and~(\ref{dim_phi_eom})
of the form
\be 
\left(\matrix{u\cr \phi\cr}\right) = \left(\matrix{u_\ast\cr \phi_\ast\cr}\right)\exp(i{\bm k}\cdot{\bm x}+\sigma t),
\ee 
where ${\bm k}\equiv k_x\hat x + k_y\hat y$, ${\bm x}\equiv x\hat x + y\hat y$ and $u_\ast$ and $\phi_\ast$ are constants.  $\Re\sigma=\Re\sigma(\bm{k})$ gives the rate with which the amplitude of the mode with wavevector $\bm{k}$ grows (for $\Re\sigma > 0$) or attenuates (for $\Re\sigma < 0$).  Equation (\ref{dim_u_eom}) yields
\be
\Re \sigma(\bm{k}) = -C_1 k_x^2 - C_2 k_y^2 - Dk^4.
\ee
It follows that the surface remains flat for positive $C_1$ and $C_2$.
For the linear theory being studied here, the experimentally observed ripple wavevector $\bm{k}_\ast$ is the one with the highest growth rate.  If $C_1<0$ and $C_1<C_2$, therefore, $\bm{k}_\ast$ lies along the $x$-direction and so-called parallel-mode ripples develop.  If $C_2<0$ and $C_2<C_1$, on the other hand, then $\bm{k}_\ast = k_\ast\bm{\hat y}$ and we have perpendicular-mode ripples.

Equations (\ref{C_1}) and (\ref{C_2}) show that the coefficients $C_1$ and $C_2$ are linear functions of $c_{s,0}$, the spatially-averaged surface concentration of $A$ atoms.  For $c_{s,0}=0$ or $c_{s,0}=1$, they assume the values for a target composed entirely of species $A$ or species $B$, respectively.   

Species $A$ is Au and species $B$ is Si in the experiments of Moon {\it et al.}\cite{Moon16}
For the argon ion beam they employed, a pure silicon target is stable while perpendicular-mode ripples would develop on a target consisting entirely of gold.  We will therefore assume that $C_1$ and $C_2$ are positive for $c_{s,0}=0$, and that $C_2<0$ and $C_2<C_1$ for $c_{s,0}=1$.
This leaves four possibilities to be considered:
\begin{enumerate}
\item{$C_1<C_2$ for $c_{s,0}=0$ and $C_1 < 0$ for $c_{s,0}=1$;}
\item{$C_1>C_2$ for $c_{s,0}=0$ and $C_1 < 0$ for $c_{s,0}=1$;}
\item{$C_1<C_2$ for $c_{s,0}=0$ and $C_1 > 0$ for $c_{s,0}=1$; and}
\item{$C_1>C_2$ for $c_{s,0}=0$ and $C_1 > 0$ for $c_{s,0}=1$.}
\end{enumerate}

We will begin with Case 1.  Figure 2 shows the corresponding
dependences of $C_1$ and $C_2$ on $c_{s,0}$.  Let $X_1$ denote the value of $c_{s,0}$ where $C_1$ vanishes.  The value of $c_{s,0}$ where $C_1=C_2$ will be denoted by $X_2$.  Clearly, $X_1 < X_2$.  There are three distinct ranges of $c_{s,0}$:
\begin{enumerate}[(a)]
\item{For $0\le c_{s,0}<X_1$, both $C_1$ and $C_2$ are positive.  In this regime, the solid surface remains flat as time passes.}
\item{For $X_1< c_{s,0} < X_2$, we have $C_1<0$ and $C_2>C_1$.  Consequently, parallel-mode ripples develop.}
\item{For $X_2 < c_{s,0} \le 1$, the inequalities $C_2 < 0$ and $C_2 < C_1$ are satisfied.  Perpendicular-mode ripples form in this case.}
\end{enumerate}
We see that if the surface concentration of impurities $c_{s,0}$ is zero, the surface remains flat as time passes.  As $c_{s,0}$ is increased, the surface stays flat until
$c_{s,0}$ passes through the critical value $X_1$. Parallel-mode ripples then emerge.  Finally, perpendicular-mode ripples form once $c_{s,0}$ has passed through the second critical value $X_2$.
We conclude that our model gives precisely the same morphological transitions as Moon {\it et al.}~observed provided that $C_1<C_2$ for $c_{s,0}=0$ and $C_1 < 0$ for $c_{s,0}=1$. 

Parallel analyses readily show that for Cases 2 - 4, the morphology switches directly from a flat state to perpendicular-mode ripples as $c_{s,0}$ increases from 0 to 1.  Thus, our model only exhibits the behavior observed by Moon and co-workers if $C_1$ and $C_2$ satisfy the two inequalities that define Case 1.

\vspace{15pt}

\centerline{\includegraphics[width=3.5in]{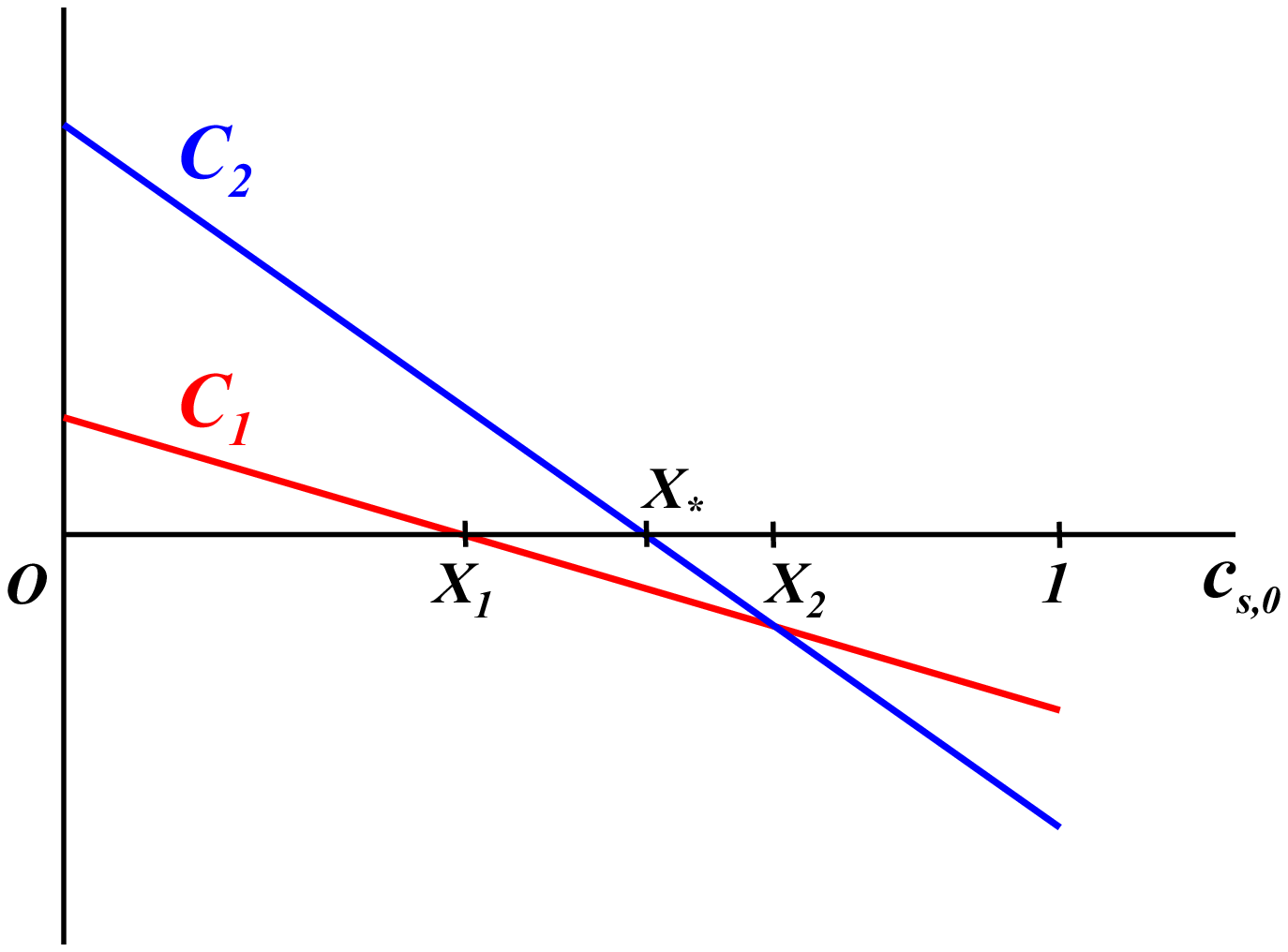}}
\begin{figure}[h]
\caption{(Color online) The curvature coefficients $C_1$ and $C_2$ plotted versus $c_{s,0}$ (the spatially-averaged surface concentration of $A$ atoms) for Case 1.}
\label{fig1}
\end{figure}

Interestingly, Moon {\it et al.}~found that for a range of surface concentrations $c_{s,0}$, the ridges of the parallel-mode ripples were sectioned into small pieces.  As they noted, this suggests that although the dominant instability was in the $x$-direction, the surface was nonetheless also unstable in the $y$-direction.  In our notation, this means that $C_1<C_2<0$.  Figure 2 shows that for Case 1, there is just such a regime: it is the regime in which $X_* < c_{s,0} < X_2$, where $X_*$ is the surface concentration where $C_2$ vanishes.

Moon {\it et al.}~observed that for gold concentrations high enough for ripples to form, the spatial modulation of the surface height can be accompanied by a spatial modulation of the surface composition.  This too can be explained by our theory.  As Eq.~(\ref{dim_phi_eom}) shows, a spatial oscillation of the surface height $h=h_0-v_0t +u$ produces a corresponding spatial oscillation of the surface composition $c_s = c_{s,0}+\phi$.  

Moon and coworkers did not find compositional variations on the surface ripples if the spatially-averaged gold concentration was too low.  At first blush, this seems to be at variance with our theory.
However, as noted by Moon and coworkers, the amplitude of the spatial oscillation of the surface composition may simply have been too small for it to have been detected.  

\section{Discussion}
\label{sec:discussion}

We have shown that once the existing theory of concurrent ion sputtering and impurity deposition has been appropriately modified, it is able to reproduce the key experimental observations made by Moon and co-workers.\cite{Moon16}  It particular, our theory gives transitions from a flat state to parallel-mode ripples and then to perpendicular-mode ripples as the spatially-averaged impurity concentration is increased. 
The required modification is to alter the Sigmund model of ion sputtering to take into account the effect that impurities have on the collision cascades.  

The Sigmund model for sputtering of elemental materials\cite{Sigmund73} has limitations, as recent simulations have revealed.\cite{Hossain12,HoblerXX}  The generalized version of that model that was introduced and used in this paper is likely no more accurate than the original model.  However, it is a valuable starting point for theoretical investigations.  In the future, atomistic simulations combined with the crater function formalism of Ref.~\onlinecite{Harrison15} could lead to a more accurate theory.

It was not necessary to incorporate the effects of phase separation into our theory in order for it to yield the experimentally-observed morphological transitions.  However, for relatively high gold concentrations, Moon {\it et al.}~observed crystalline gold nanoparticles embedded in a matrix of amorphous silicon that presumably had some unagglomerated gold interspersed within it.  Our work suggests that phase separation plays a relatively minor role in the experiments of Moon {\it et al.}, but it would likely have to be included to make a theory that gives detailed agreement with their experiments.

For the sake of simplicity, in our theory we assumed that thermally activated surface diffusion is negligible compared to ion-induced surface viscous flow.  This is likely a good approximation in the experiments of Moon {\it et al.} except in sample regions where the gold concentration was very high.  The approximation fails in those regions because ion-induced viscous flow does not occur in metallic solids.

Although simplifications were made in formulating our theory, it does represent a considerable improvement on the model Moon {\it et al.}~advanced in an attempt to explain their experimental results.\cite{Moon16}  In their theory, Moon {\it et al.}~took the incident ion beam and the impurities to be normally incident on the sample surface even though the ions and impurities were incident at near grazing angles in their experiments. They also neglected the effect that subsurface impurities have on the collision cascades.  The resulting model does not account for the formation of ripples or for the switch in the direction of the predominant ripple wavevector as the gold concentration increases.

The experiments of Moon and co-workers demonstrate that impurity co-deposition can induce the formation of nanostructures on silicon during ion bombardment even if silicide formation does not occur.  This finding runs counter to the view that silicide formation is either a precondition for pattern formation or is at least the dominant driving force behind it.\cite{Zhang11,Zhang12,Hofsass13,Engler14}  However, it is consistent with prior theoretical work that showed that although silicide formation promotes the emergence of nanostructures,\cite{Bradley13} it is not essential for it to occur.\cite{Bradley11imp,Bradley12}  Moreover, in experimental work that predates that of Moon {\it et al.}, a broad gold ion beam incident on a silicon surface produced nanostructures.\cite{Mollick14}  This too gives evidence that impurities can induce pattern formation on silicon even if silicides are not formed.

\section{Conclusions}
\label{sec:conclusions}

If the surface of an elemental material remains flat when it is bombarded with a broad beam of noble gas ions, in some cases the surface can be destabilized by the co-deposition of impurities.  Nanoscale surface patterns are the result.  Important recent experiments carried out by Moon and co-workers show that the surface of silicon can be destabilized by the co-deposition of gold, and so demonstrate that an instability can be induced even if silicide formation does not occur.\cite{Moon16}  This runs counter to the view that for co-deposition to destabilize a silicon surface, silicide formation must take place. Moon {\it et al.}~also observed a sequence of morphological transitions as the spatially-averaged gold concentration is increased that cannot be explained by the previously advanced theory.

In this paper, we generalized the Sigmund model of ion sputtering to take into account the effect that near-surface impurities have on the collision cascades. This generalization was then used to refine the theory of ion bombardment with concurrent impurity deposition.
With this modification, as the spatially-averaged impurity concentration is increased, the theory yields a flat surface, parallel-mode ripples, and finally perpendicular-mode ripples in succession for a certain range of the parameters.  This is precisely the sequence of morphological transitions that
Moon {\it et al.}~observed.

\begin{acknowledgments}

R.M.B. is grateful to the National Science Foundation for its support through grant DMR-1305449.

\end{acknowledgments}

\vfill\eject


\begin{thebibliography}{00}

\bibitem{Ziberi09} B. Ziberi, M. Cornejo, F. Frost, and B. Rauschenbach, 
J. Phys. Condens. Matter {\bf 21}, 224003 (2009).

\bibitem{Facsko99}  S. Facsko, T. Dekorsy, C. Koerdt, C. Trappe, H. Kurz, A. Vogt, and H. L. Hartnagel, Science {\bf 285}, 1551 (1999).

\bibitem{Frost00} F. Frost, A. Schindler, and F. Bigl, Phys. Rev. Lett. {\bf 85}, 4116 (2000).

\bibitem{Munoz-Garcia14}
J. Mu\~{n}oz-Garc\'{i}a, L. V\'{a}zquez, M. Castro, R. Gago, A. Redondo-Cubero, A. Moreno-Barrado and R. Cuerno, Mater. Sci. Eng. R-Rep. {\bf 86}, 1 (2014).

\bibitem{Ozaydin05} G. Ozaydin, A. S. \"Ozcan, Y. Wang, K. F. Ludwig, H. Zhou, R. L. Headrick, and D. P. Siddons, Appl. Phys. Lett. {\bf 87}, 163104 (2005).

\bibitem{Ozaydin08} G. Ozaydin-Ince, K. F. Ludwig, Jr., H. Zhou, and R. L. Headrick, J. Vac. Sci. Technol. B {\bf 26}, 551 (2008).

\bibitem{Hofsass08} H. Hofs\"ass and K. Zhang, Appl. Phys. A: Mater. Sci. Process. {\bf 92}, 517 (2008).

\bibitem{Sanchez-Garcia08} J. A. S\'anchez-Garc\'ia, L. V\'azquez, R. Gago, A. Redondo-Cubero,
J. M. Albella, and Zs. Czigany, Nanotechnology {\bf 19}, 355306 (2008).

\bibitem{Ozaydin-Ince09} G. Ozaydin-Ince and K. F. Ludwig, Jr., J. Phys.: Condens. Matt. {\bf 21}, 224008 (2009).

\bibitem{Macko10} S. Macko, F. Frost, B. Ziberi, D. F. F\"orster, and T. Michely, Nanotechnology {\bf 21}, 085301 (2010).

\bibitem{Zhang10} K. Zhang, H. Hofs\"ass, and H. Zutz, Nuclear Inst. Meth. Phys. Res. B {\bf 268}, 1967 (2010).

\bibitem{Zhou11} J. Zhou, S. Facsko, M. Lu, and W. M\"oller, J. Appl. Phys. {\bf 109}, 104315 (2011).

\bibitem{Cornejo11} M. Cornejo, B. Ziberi, C. Meinecke, D. Hirsch, J. W. Gerlach, T. H\"oche, F. Frost, and B. Rauschenbach, 
Appl. Phys. A {\bf 102}, 593 (2011).

\bibitem{Zhang11} K. Zhang, M. Br\"otzmann, and H. Hofs\"ass, New J. Phys. {\bf 13}, 013033 (2011).

\bibitem{Macko11} S. Macko, F. Frost, M. Engler, D. Hirsch, T. H\"oche, J. Grenzer, and T. Michely, 
New Journal of Physics {\bf 13}, 073017 (2011).

\bibitem{Hofsass12} H. Hofs\"ass, K. Zhang, A. Pape, O. Bobes, and M. Br\"otzmann, Appl. Phys. A  (2012) doi: 10.1007/s00339-012-7285-8.

\bibitem{Macko12} S. Macko, J. Grenzer, F. Frost, M. Engler, D. Hirsch, M. Fritzsche, A. M\"ucklich, and T. Michely,
New Journal of Physics {\bf 14}, 073003 (2012).

\bibitem{Zhang12} K. Zhang, M. Br\"otzmann, and H. Hofs\"ass, AIP Advances {\bf 2}, 032123 (2012).

\bibitem{Redondo-Cubero12} A. Redondo-Cubero, R. Gago, F. J. Palomares, A. M\"ucklich, M. Vinnichenko, and L. V\'azquez,
Phys. Rev. B {\bf 86}, 085436 (2012).

\bibitem{Hofsass13} H. Hofs\"ass, K. Zhang, A. Pape, O. Bobes, and M. Br\"otzmann, Appl. Phys. A {\bf 111}, 653 (2013).

\bibitem{Bhattacharjee13} S. Bhattacharjee, P. Karmakar, V. Naik, A. K. Sinha, and A. Chakrabarti,
Appl. Phys. Lett. {\bf 103}, 181601 (2013).

\bibitem{Khanbabaee14} B. Khanbabaee, D. L\"utzenkirchen-Hecht, R. H\"ubner, J. Grenzer, S. Facsko, and U. Pietsch, J. Appl. Phys. {\bf 116}, 024301 (2014).

\bibitem{Engler14} M. Engler, F. Frost, S. M\"uller, S. Macko, M. Will,
R. Feder, D. Spemann, R. H\"ubner, S. Facsko, and T. Michely, Nanotechnology {\bf 25}, 115303 (2014).

\bibitem{Gago14} R. Gago, A. Redondo-Cubero, F. J. Palomares, and L. V\'azquez,
Nanotechnology {\bf 25}, 415301 (2014).

\bibitem{Ko15} T.-J. Ko, K. H. Oh, and M.-W. Moon, Adv. Mater. Interfaces {\bf 2}, 1400431 (2015). 

\bibitem{Vayalil15} S. K. Vayalil, A. Gupta, S. V. Roth, and V. Ganesan, J. Appl. Phys. {\bf 117}, 024309 (2015).


\bibitem{Bradley12} R. M. Bradley, Phys. Rev. B {\bf 85}, 115419 (2012).

\bibitem{Bradley13} R. M. Bradley, Phys.~Rev.~B {\bf 87}, 205408 (2013).

\bibitem{Moon16} B. Moon, S. Yoo, J.-S. Kim, S. J. Kang, J. Munoz-Garc\'ia, and R. Cuerno, arXiv:1601.02534 (2016).

\bibitem{Sigmund73} P. Sigmund, J. Mater. Sci. {\bf 8}, 1545 (1973).

\bibitem{Bradley11exact} R. M. Bradley, Phys. Rev. B {\bf 84}, 075413 (2011).

\bibitem{Shenoy07} V. B. Shenoy, W. L. Chan, and E. Chason, Phys. Rev. Lett. {\bf 98}, 256101 (2007).

\bibitem{Bradley10} R. M. Bradley and P. D. Shipman, Phys. Rev. Lett. {\bf 105}, 145501 (2010).

\bibitem{Shipman11} P. D. Shipman and R. M. Bradley, Phys. Rev. B {\bf 84}, 085420 (2011).

\bibitem{Bradley11imp} R. M. Bradley, Phys. Rev. B {\bf 83}, 195410 (2011).

\bibitem{Bradley12ASS} R. M. Bradley and P. D. Shipman, Appl. Surf. Sci. {\bf 258}, 4161 (2012).

\bibitem{Bradley88} R. M. Bradley and J. M. E. Harper, J. Vac. Sci. Technol. A {\bf 6}, 2390 (1988).

\bibitem{Motta14} F. C. Motta, P. D. Shipman, and R. M. Bradley, Phys.~Rev.~B {\bf 90}, 085428 (2014). 

\bibitem{Umbach01} C. C. Umbach, R. L. Headrick, and K.-C. Chang, Phys. Rev. Lett. {\bf 87}, 246104 (2001).

\bibitem{Orchard63} S. E. Orchard, Appl. Sci. Res. {\bf 11A}, 451 (1963).

\bibitem{Carter96} G. Carter and V. Vishnyakov, Phys. Rev. B {\bf 54}, 17647 (1996).

\bibitem{Moseler05} M. Moseler, P. Gumbsch, C. Casiraghi, A. C. Ferrari, and J. Robertson, Science {\bf 309}, 1545 (2005).

\bibitem{Davidovitch07} B. Davidovitch, M. J. Aziz, and M. P. Brenner, Phys. Rev. B {\bf 76}, 205420 (2007).

\bibitem{Hossain12} M.~Z. Hossain, J.~B. Freund, and H.~T. Johnson, J. Appl. Phys. {\bf 111}, 103513 (2012).

\bibitem{HoblerXX} G. Hobler, R. M. Bradley, and H.~M.~Urbassek,  submitted to Phys. Rev. B.

\bibitem{Harrison15} M. P. Harrison and R. M. Bradley,  J.~Phys.:~Cond.~Matt.~{\bf 27}, 295301 (2015).

\bibitem{Mollick14} S. A. Mollick, D. Ghose, P. D. Shipman, and R. M. Bradley, Appl.~Phys.~Lett.~{\bf 104}, 043103 (2014).









\end{thebibliography}
\end{document}